\documentclass[12pt]{article}
\usepackage{graphicx}

\newcommand{\beq}{\begin{equation}}
\newcommand{\eeq}{\end{equation}}
\newcommand{\beqn}{\begin{eqnarray}}
\newcommand{\eeqn}{\end{eqnarray}}
\newcommand{\beqns}{\begin{eqnarray*}}
\newcommand{\eeqns}{\end{eqnarray*}}

\begin{document}

\begin{titlepage}
\begin{center}

\hfill USTC-ICTS-18-19\\
\hfill November 2018

\vspace{2.5cm}

{\large {\bf Note on invisible decays of light mesons}}\\
\vspace*{1.0cm}
 {Dao-Neng Gao$^\dagger$ \vspace*{0.3cm} \\
{\it\small Interdisciplinary Center for Theoretical Study,
University of Science and Technology of China, Hefei, Anhui 230026
China}}

\vspace*{1cm}
\end{center}
\begin{abstract}
\noindent
A search for the invisible decays of $\omega$ and $\phi$ mesons in $J/\psi\to \omega (\phi)\eta$ transitions has been performed by the BESIII Collaboration very recently. Inspired by this experimental study, we compute the lowest order contribution to branching ratios of ${\cal B}(V\to\bar{\nu}\nu)$ with $V$ denoting $\rho,\;\omega,\;\phi$, as the standard model background to these invisible decays. Our predictions are far below the upper bounds given by the BESIII experiment. We also analyze the $J/\psi\to \eta(\eta^\prime)\bar{\nu}\nu$ processes, and estimate their decay rates. Furthermore, the invisible decays of light pseudoscalar mesons $P$ including $\pi^0$, $\eta$, and $\eta^\prime$ are reexamined in the present note. It is shown that, due to the helicity suppression of the two-neutrino final state, the standard model contributions to $P\to{\rm invisible}$ decays are dominated by $P\to\bar{\nu}\nu \bar{\nu}\nu$ processes.
\end{abstract}

\vfill
\noindent
$^{\dagger}$ E-mail:~gaodn@ustc.edu.cn
\end{titlepage}

The study of quarkonium decays to invisible final states is an interesting topic both theoretically and experimentally. The BESII \cite{BESII2008} and $BABAR$ \cite{Babar2009} experiments have explored the invisible decays of heavy quarkonia including $J/\psi$ and $\Upsilon(1S)$, and some interesting upper limits on the decay rates are obtained. These limits are still above the standard model (SM) predictions \cite{CLN1998}.
For the light mesons like pseudoscalar $P$ with $P$ denoting $\pi^0,\;\eta,\;\eta^\prime$, the current experimental upper bounds have been given by the E949 Collaboration \cite{E9492005} for $\pi^0\to {\rm invisible}$, and by the BESIII Collaboration \cite{BESIII2013} for $\eta (\eta^\prime)\to {\rm invisible}$ decays; while branching ratios of these pseudoscalar invisible decays were calculated in Ref. \cite{AMP1982} long time ago.

Very recently, using a data sample of $(1310.6\pm 7.0)\times 10^6$ $J/\psi$ events, the first experimental search for invisible decays of a light vector meson $V$ has been performed by the BESIII Collaboration \cite{BESIII2018} via $J/\psi\to V\eta$ ($V=\omega,\;\phi$) decays, and the upper limits on the ratio have been measured to be
\beq\label{upperboundomega}\frac{{\cal B}(\omega\to {\rm invisible})}{{\cal B}(\omega\to \pi^+\pi^-\pi^0)}< 8.1\times 10^{-5}\eeq
and
\beq\label{upperboundphi}\frac{{\cal B}(\phi\to {\rm invisible})}{{\cal B}(\phi\to K^+ K^-)}< 3.4 \times 10^{-4}, \eeq
at the $90\%$ confidence level, for $\omega$ and $\phi$ mesons, respectively. Using the branching ratios of ${\cal B}(\omega\to\pi^+\pi^-\pi^0)$ and ${\cal B}(\phi\to K^+ K^-)$ by the Particle Data Group (PDG) \cite{PDG2018}, we get
 \beq\label{omegabound}{\cal B}(\omega\to {\rm invisible})<7.3\times 10^{-5}\eeq
 and \beq\label{phibound}{\cal B}(\phi\to {\rm invisible})<1.7\times 10^{-4}.
\eeq
 In general, these invisible final states can be neutrinos in the SM, and also some new particles beyond the SM, which, for instance, could be the candidate of light dark matter, as discussed in Refs. \cite{McElrath2007, BP2010}. The decays $V\to \chi\chi$ with $\chi$ denoting the light dark matter particles have been estimated, and the branching ratio is predicted to be up to the order of $10^{-8}$ \cite{McElrath2007}. The authors of Refs. \cite{BP2010, BFG2018} have analyzed invisible decays of heavy mesons $B(D)$ and strange meson $K_L$, in order to search for light dark matter particles in these processes.  Thus, the investigation of invisible meson decays may help us to explore the novel dynamics or impose useful constraints on some models beyond the SM.

The first motivation of the present note is to study the invisible decays of light vector mesons including $\rho$, $\omega$, and $\phi$ in the SM. Since neutrinos are the only invisible particles in the SM, we will focus on the analysis of the neutrino contributions for these processes. It is expected that these contributions should be very small, however, a quantitative analysis, to our knowledge, has not been done yet. Only after we fully understand the SM background can the future precise experimental investigations of the invisible decays possibly provide us with some interesting information on new physics.

One can easily find that the invisible decays of $V$ in the SM, $V\to \bar{\nu} \nu$, are given by the neutral current interactions, which can be expressed as
\beq\label{NC}
{\cal L}_{\rm NC}=e J_\mu^{\rm em} A^\mu+\frac{g}{\cos\theta_W}J^Z_\mu Z^\mu\eeq
with
\beq\label{emcurrent}
J_\mu^{\rm em}=\sum_f Q_f\bar{f}\gamma_\mu f,
\eeq
and
\beq\label{weakneutralcurrent}
J_\mu^Z=\frac{1}{2}\sum_f \bar{f}\gamma_\mu(g_V^f-g_A^f\gamma_5)f,\eeq
 where $e$ is the coupling constant of electromagnetic interaction, $g$ is the SU(2)$_L$ coupling constant, $\theta_W$ is the Weinberg angle, and $f$ denotes fermions including leptons and quarks.  Also $g_V^f=T_3^f-2 Q_f \sin^2 \theta_W$ and $g_A^f=T_3^f$,  where $Q_f$ is the charge, and $T_3^f$ is the third component of the weak isospin of the fermion.

\begin{figure}[t]
\begin{center}
\includegraphics[width=5cm,height=2cm]{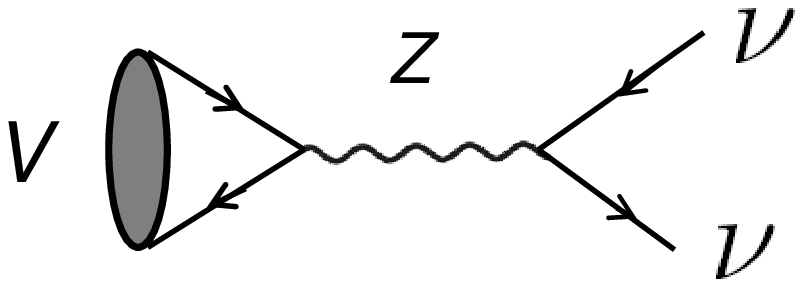}
\end{center}
\caption{Lowest-order diagrams for the decay $V \to \bar{\nu}{\nu}$. For $V\to e^+e^-$ decays,  replace $Z$ with the photon $\gamma$, and the final states will be the charged leptons.}\label{figure1}
\end{figure}

\begin{table}[t]\begin{center}\begin{tabular}{ c|c| c | c} \hline\hline
 Meson $V$ & $G_{V}$& $Q_{V}$ & $R_V$\\\hline
 $\rho$ & $\frac{1}{\sqrt{2}}\left(\frac{1}{2}-\sin^2\theta_W\right)$& $\frac{1}{\sqrt{2}}$ & $5.10\times 10^{-9}$\\\hline
$\omega$& $\frac{-\sin^2\theta_W}{3\sqrt{2}}$ & $\frac{1}{3\sqrt{2}}$ &$3.84\times 10^{-9}$ \\\hline
$\phi$& $-\frac{1}{4}+\frac{1}{3}\sin^2\theta_W$& $-\frac{1}{3}$& $5.65\times 10^{-8}$\\\hline
\hline
\end{tabular}\caption{Ratios of the decay rates $R_V$ for $\rho$, $\omega$, and $\phi$ mesons.} \end{center}\end{table}

For $V\to \bar{\nu}{\nu}$ decays, only the intermediate $Z$ boson can contribute, which has been shown in Fig. \ref{figure1}. For the charged leptonic decays, $V\to e^+ e^-$, both the photon and $Z$ contribute; however, contributions from $Z$ boson could be negligible. Thus, direct calculations will give the ratio as
\beq\label{ratio1}
R_V\equiv\frac{{\cal B}(V\to \bar{\nu} \nu)}{{\cal B}(V\to e^+ e^-)}=\frac{G_F^2m_V^4}{4\pi^2\alpha^2_{\rm em}}\frac{G_V^2}{Q_V^2}\cdot 3
\eeq
for $V=\rho$, $\omega$, and $\phi$, where the factor 3 in the equation is due to the neutrino flavors, $\alpha_{\rm em}=e^2/4\pi$, and $G_F$ is Fermi coupling constant with
\beq\label{GF}\frac{G_F}{\sqrt{2}}=\frac{g^2}{8m_W^2}=\frac{g^2}{8 m_Z^2\cos^2\theta_W}.\eeq
 $G_V$ and $Q_V$ have been listed in Table 1, and the values of $R_V$ have been calculated.  By employing the branching ratios of $V\to e^+ e^-$ from Ref. \cite{PDG2018}, one can get
 \beqn\label{br-rho}
 {\cal B}(\rho\to \bar{\nu}{\nu})=(2.41\pm 0.02)\times 10^{-13},\\\label{br-omega}
 {\cal B}(\omega\to \bar{\nu}{\nu})=(2.79\pm 0.05)\times 10^{-13},\\\label{br-phi}
 {\cal B}(\phi\to \bar{\nu}{\nu})=(1.67\pm 0.02)\times 10^{-11}.
 \eeqn
 Here the uncertainty is due to the experimental value of ${\cal B}(V\to e^+ e^-)$ only. These are quite small branching ratios, which means a big experimental challenge. Meanwhile, comparing with the present experimental bound in eqs. (\ref{omegabound}) and (\ref{phibound}) reported by the BESIII Collaboration, and theoretical predictions ${\cal B}(V\to\chi\chi)\sim 10^{-8}$ given in Ref. \cite{McElrath2007}, some interesting room for new physics in these invisible decays might be expected.

Note that this experimental search at BESIII \cite{BESIII2018} was performed via $J/\psi \to V\eta$ ($V=\omega,\;\phi$) decays. By taking the current experimental measurements of $J/\psi\to V\eta$ by PDG \cite{PDG2018}, together with our predictions, we obtain
\beqn\label{br-rhonunu}
{\cal B}(J/\psi\to \eta \rho\to \eta\bar{\nu}\nu)={\cal B}(J/\psi\to \eta \rho)\cdot{\cal B}(\rho\to \bar{\nu}\nu)=(4.65\pm 0.55)\times 10^{-17},\\\label{br-omeganunu}
{\cal B}(J/\psi\to \eta \omega\to \eta\bar{\nu}\nu)={\cal B}(J/\psi\to \eta \omega)\cdot{\cal B}(\omega\to \bar{\nu}\nu)=(4.85\pm 0.56)\times 10^{-16},\\\label{br-phinunu}
{\cal B}(J/\psi\to \eta \phi\to \eta\bar{\nu}\nu)={\cal B}(J/\psi\to \eta \phi)\cdot{\cal B}(\phi\to \bar{\nu}\nu)=(1.25\pm 0.14)\times 10^{-14}.
\eeqn
This indicates that, on the other hand, it will be of interest to investigate the $J/\psi\to \eta \bar{\nu}\nu$ decay in which $\bar{\nu}\nu$ is not from any resonances, and the analysis may provide some complementary information for the future experimental study.

In the SM, it is natural to believe that, at the leading order, this process proceeds through $J/\psi\to Z^*\eta$, followed by $Z^*\to \bar{\nu}\nu$. Considering charge conjugate invariance in $J/\psi\to Z^*\eta$ transition, one can effectively write
\beq\label{JZeta}
{\cal L}_{JZ\eta}=\frac{g_{JZ\eta}}{m_{J}}\varepsilon^{\mu\nu\alpha\beta}J_{\mu\nu}Z_{\alpha\beta}\eta,
\eeq
where $J_{\mu\nu}=\partial_\mu J_\nu-\partial_\nu J_\mu$, $Z_{\alpha\beta}=\partial_\alpha Z_\beta-\partial_\beta Z_\alpha$, and $g_{JZ\eta}$ is an unknown dimensionless coupling. Thus, the differential decay rate of the transition $J/\psi\to \eta Z^*\to \eta\bar{\nu}\nu$ is expressed as
\beq\label{diff-rate}
\frac{d\Gamma(J/\psi\to \eta\bar{\nu}\nu)}{d q^2}=\frac{G_F^2 g^2_{JZ\eta}q^2}{144\pi^3 m_J^5}\lambda^{3/2}(m_J^2,m_\eta^2, q^2),
\eeq
where $q^2$ is the neutrino pair invariant mass squared with its range $0 \leq q^2\leq (m_J-m_\eta)^2$, and $\lambda(a,b,c)=a^2+b^2+c^2- 2 ab-2 bc- 2 ac$. Consequently, we have
\beq\label{rate}\Gamma(J/\psi\to \eta\bar{\nu}\nu)=\frac{G_F^2 m_J^5 g^2_{JZ\eta}}{144\pi^3} f(r_\eta) \cdot 3
\eeq
with $r_\eta=m_\eta^2/m_J^2$, $f(x)=(1-x)(x^4-14 x^3-94 x^2-14 x+1)/20 -3 x^2 (1+x)\log x$, and the factor 3 is also due to the neutrino flavors.  In order to predict this decay rate, one has to fix the unknown coupling $g_{JZ\eta}$.

\begin{figure}[t]
\begin{center}
\includegraphics[width=5.5cm,height=2cm]{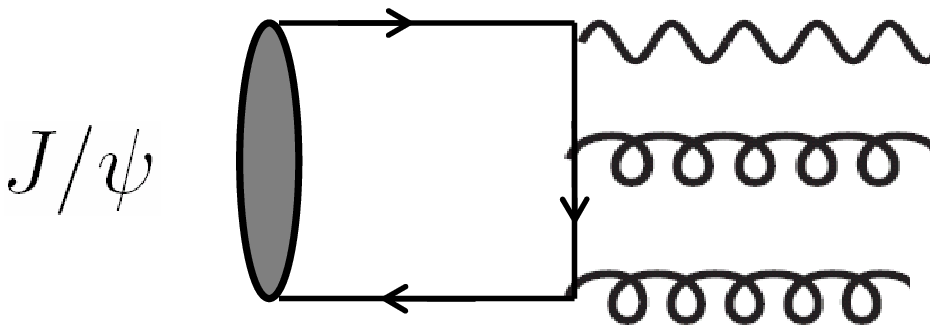}
\end{center}
\caption{Representative diagrams for $J/\psi\to \gamma/Z + gg$ transitions with the wave line denoting $\gamma$ or $Z$, and the curly line denoting the gluon. By exchanging the $\gamma/Z$ line with the gluon line, one can totally get three diagrams.}\label{figure2}
\end{figure}

\begin{figure}[t]
\begin{center}
\includegraphics[width=11cm,height=8cm]{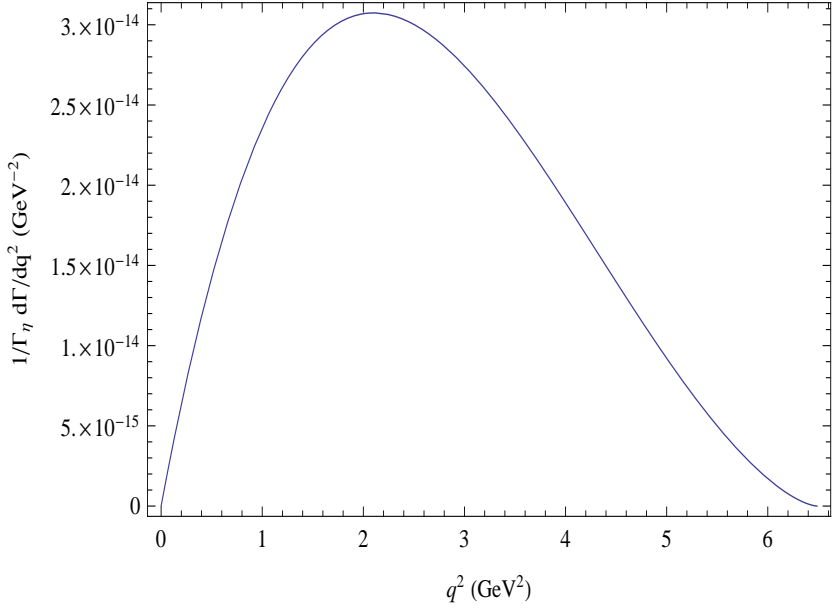}
\end{center}
\caption{The decay spectrum for $J/\psi \to \eta\bar{\nu}\nu$ normalized by $\Gamma(J/\psi\to \eta\gamma)$ denoted by $\Gamma_\eta$ as the function of the neutrino pair invariant mass squared $q^2$.}\label{figure3}
\end{figure}

\begin{figure}[t]
\begin{center}
\includegraphics[width=11cm,height=8cm]{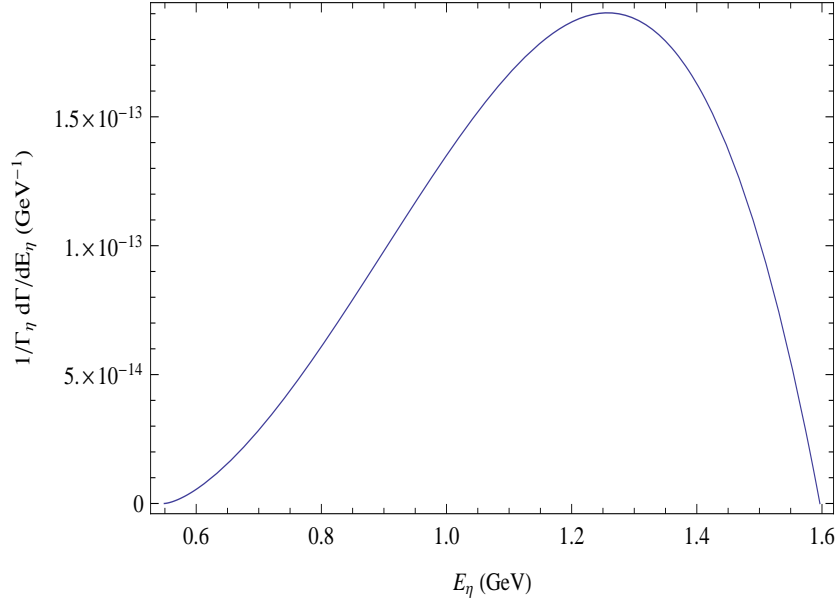}
\end{center}
\caption{The decay spectrum for $J/\psi \to \eta\bar{\nu}\nu$ normalized by $\Gamma(J/\psi\to \eta\gamma)$ denoted by $\Gamma_\eta$ as the function of the energy of $\eta$ meson $E_\eta$.}\label{figure4}
\end{figure}

It is assumed that the radiative $J/\psi$ decays like $J/\psi\to \gamma \eta(\eta^\prime)$, which are suppressed by the Okubo-Zweig-Iizuka rule, are dominated by the process $J/\psi\to \gamma gg$ with $gg\to \eta(\eta^\prime)$ \cite{NSVZ1980,KKKS1983, YLSZ1984}. It is expected that $J/\psi\to Z^*\eta (\eta^\prime)$ transitions can be described in the same way, and the corresponding diagrams have been displayed in Fig. \ref{figure2}. One can parameterize the effective vertex of radiative decays as
\beq\label{JGammaeta}
{\cal L}_{J\gamma\eta}=\frac{g_{J\gamma\eta}}{m_{J}}\varepsilon^{\mu\nu\alpha\beta}J_{\mu\nu}F_{\alpha\beta}\eta,
\eeq
where $g_{J\gamma\eta}$ is an effective coupling. Due to charge conjugate invariance, the leading order contribution to $J/\psi\to Z^*\eta$ is only given by the vector part of $J_\mu^Z$ in eq. (\ref{weakneutralcurrent}). Then it is easy from Fig. \ref{figure2} to see
\beq\label{couplingratio}
\frac{g_{JZ\eta}}{g_{J\gamma\eta}}=\frac{g g_V^c/2 \cos\theta_W}{e Q_C}=\frac{3/8-\sin^2\theta_W}{\cos\theta_W\sin\theta_W}.\eeq
This relation would help us to estimate the decay rate of $J/\psi\to \eta\bar{\nu}\nu$ using the experimental information of $J/\psi\to \eta\gamma$ decay. It is thus straightforward to get
\beq\label{ratio:br}
R_{\bar{\nu}\nu}^\eta\equiv\frac{{\cal B}(J/\psi\to \eta\bar{\nu}\nu)}{{\cal B}(J/\psi\to \eta\gamma)}=\frac{G_F^2 m_J^4}{16\pi^2}\frac{f(r_\eta)}{(1-r_\eta)^3}\left(\frac{g_{JZ\eta}}{g_{J\gamma\eta}}\right)^2 =3.4\times 10^{-13}.
\eeq
Accordingly, taking ${\cal B}(J/\psi\to \eta\gamma)=(1.104\pm 0.034)\times 10^{-3}$ from Ref. \cite{PDG2018}, we have
\beq\label{br:jpsinunueta}
{\cal B}(J/\psi\to \eta\bar{\nu}\nu)=(3.8\pm 0.1)\times 10^{-16},
\eeq
which can be compared with the results in eqs. (\ref{br-rhonunu}), (\ref{br-omeganunu}), and (\ref{br-phinunu}).
Also, the decay spectrum normalized by $\Gamma(J/\psi\to \eta\gamma)$ has been plotted, as the function of the neutrino pair invariant mass squared $q^2$ in Fig. \ref{figure3} and as the function of the energy of $\eta$ meson $E_\eta$ in Fig. \ref{figure4}.

Similar analysis can be applied to the $J/\psi\to \eta^\prime\bar{\nu}\nu$ decay. Replacing $\eta$ in eq. (\ref{ratio:br}) by $\eta^\prime$, one has
 \beq\label{ratioetaprime:br}
R_{\bar{\nu}\nu}^{\eta^\prime}\equiv\frac{{\cal B}(J/\psi\to \eta^\prime\bar{\nu}\nu)}{{\cal B}(J/\psi\to \eta^\prime\gamma)}=2.0\times 10^{-13},
\eeq
which, using the data ${\cal B}(J/\psi\to \eta^\prime\gamma)=(5.13\pm 0.17)\times 10^{-3}$ given in \cite{PDG2018}, gives
\beq\label{br:jpsinunuetaprime}
{\cal B}(J/\psi\to \eta^\prime\bar{\nu}\nu)=(10.2\pm 0.3)\times 10^{-16}.
\eeq
This can be compared with
\beqn\label{br-prime-rhonunu}
{\cal B}(J/\psi\to \eta^\prime \rho\to \eta^\prime\bar{\nu}\nu)={\cal B}(J/\psi\to \eta^\prime \rho)\cdot{\cal B}(\rho\to \bar{\nu}\nu)=(1.95\pm 0.19)\times 10^{-17},\\\label{br-prime-omeganunu}
{\cal B}(J/\psi\to \eta^\prime \omega\to \eta^\prime\bar{\nu}\nu)={\cal B}(J/\psi\to \eta^\prime \omega)\cdot{\cal B}(\omega\to \bar{\nu}\nu)=(5.27\pm 0.51)\times 10^{-17},\\\label{br-prime-phinunu}
{\cal B}(J/\psi\to \eta^\prime \phi\to \eta^\prime\bar{\nu}\nu)={\cal B}(J/\psi\to \eta^\prime \phi)\cdot{\cal B}(\phi\to \bar{\nu}\nu)=(7.68\pm 0.84)\times 10^{-15},
\eeqn
where we have combined the experimental data of $J/\psi\to V\eta^\prime$ for $V=\rho,\;\omega,\; \phi$ by the PDG \cite{PDG2018} with our predictions in eqs. (\ref{br-rho}), (\ref{br-omega}), and (\ref{br-phi}).

Our theoretical estimations in eqs. (\ref{br:jpsinunueta}) and (\ref{br:jpsinunuetaprime}) also indicate that it is interesting to search for $J/\psi\to \eta(\eta^\prime)+ {\rm invisible}$ decays to explore new physics beyond the SM, by using very huge $J/\psi$ samples at BESIII. Actually, some model studies, for instance, $J/\psi\to P \gamma^\prime$ with $\gamma^\prime$ called as a dark photon, which is a singlet under the SM gauge groups but couples to the SM photon via kinetic mixing \cite{Holdom1986}, have been performed in \cite{BESIII2018-2} very recently. Theoretical analysis was done in \cite{Liyang2012}.

The remainder of the present note is to revisit the invisible decays of light pseudoscalar mesons including $\pi^0$, $\eta$, and $\eta^\prime$. As mentioned above, experimentally,
\beq\label{pi-nunulimit}
{\cal B}(\pi^0\to\bar{\nu}\nu)<2.7\times 10^{-7}\eeq
has been obtained, at the $90\%$ confidence level, by the E949 Collaboration \cite{E9492005};
and the upper limits
\beq\label{eta-nunulimit}
\frac{{\cal B}(\eta\to {\rm invisible})}{{\cal B}(\eta\to \gamma\gamma)}<2.6\times 10^{-4},\eeq
and
\beq\label{etaprime-nunulimit}\frac{{\cal B}(\eta^\prime\to {\rm invisible})}{{\cal B}(\eta^\prime\to \gamma\gamma)}<2.4\times 10^{-2}
\eeq
have been determined, at the $90\%$ confidence level, by the BESIII Collaboration \cite{BESIII2013}. Recently, an interesting experimental project, by searching for invisible decays of $\pi^0$,$\eta$, $\eta^\prime$, $K_S$, and $K_L$ to probe new physics, has been proposed \cite{Gninenko2015} and is designed for the NA64 experiment at the CERN SPS. Theoretically, calculation of the $P\to\bar{\nu}\nu$ decay rate in the SM has been performed in Ref. \cite{AMP1982} more than thirty years before. Due to the helicity suppression, this mode is forbidden for the massless neutrino in the SM. Such suppression does not happen in the above case for light vector mesons. If the $Z$ boson couples to a massive neutrino with the standard weak interaction, one will get, for $\pi^0$ decaying into the neutrino pair \cite{AMP1982},
\beq\label{pi0nunu}
\Gamma(\pi^0\to\bar{\nu}\nu)=\frac{G_F^2f_\pi^2 m_\pi^3}{16\pi}r_\nu^2,\eeq
where $r_\nu=m_\nu/m_\pi$, and $f_\pi=93$ MeV is the pion decay constant. For neutrino masses, if assuming $m_\nu \sim \sum_i m_{\nu_i}<0.62$ eV given in \cite{GHMT2006}, we get
\beq\label{br-pinunu}{\cal B}(\pi^0\to\bar{\nu}\nu)\sim 6.3\times 10^{-25}.
\eeq
The authors of Ref. \cite{AMP1982} obtained a larger branching ratio for $\pi^0\to \bar{\nu}\nu$ since they used a larger bound for neutrino masses.
 Similarly, one can have ${\cal B}(\eta/\eta^{\prime}\to \bar{\nu}\nu)$, which is of the same order of magnitude as eq. (\ref{br-pinunu}) or even smaller. These predictions are far from the present experimental upper bounds shown in eqs. (\ref{pi-nunulimit}) and (\ref{eta-nunulimit}).

However, this is not the whole story for $P\to {\rm invisible}$ decays from the neutrino background. Besides $P\to \bar{\nu}\nu$, there could also exist $P\to \bar{\nu}\nu \bar{\nu}\nu$ or even more numbers of neutrino pairs.
As shown above, the decay to one neutrino pair is strongly helicity-suppressed due to the very tiny neutrino mass; while such helicity suppression can be overcome for the four-neutrino final state. In the SM, the lowest order contribution to $P\to \bar{\nu}\nu \bar{\nu}\nu$ is given by the transition $P\to Z^* Z^*$ with the virtual $Z^*\to \bar{\nu}\nu$.

It is well known that $P\to \gamma\gamma$ decays are generated from chiral anomaly, which can be expressed as
\beq\label{pgammagamma}{\cal L}_{P\gamma\gamma}\sim \frac{e^2}{16\pi^2 f_P}\varepsilon^{\mu\nu\alpha\beta}F_{\mu\nu}F_{\alpha\beta} P.\eeq
Analogously, one may get
\beq\label{PZZ}{\cal L}_{PZZ}\sim \frac{g^2}{16\pi^2 f_P\cos^2\theta_W}\varepsilon^{\mu\nu\alpha\beta}Z_{\mu\nu}Z_{\alpha\beta} P.
\eeq
Thus, using naive dimensional analysis, we obtain
\beq\label{ratio:p4nu} R_P\equiv\frac{{\cal B}(P\to \bar{\nu}\nu \bar{\nu}\nu)}{{\cal B}(P\to\gamma\gamma)}\sim\frac{G_F^4 m_P^8}{\alpha_{\rm em}^2}.\eeq
In general, some different factors should appear in eqs. (\ref{pgammagamma}) and (\ref{PZZ}) for $P=\pi^0,\;\eta,\;\eta^\prime$. However, as an order of magnitude estimation of eq. (\ref{ratio:p4nu}) in the present note, we neglect these factors. Numerically, we have $R_{\pi^0}\sim 3.8\times 10^{-23}$, $R_\eta\sim 2.9\times 10^{-18}$, and $R_{\eta^\prime}\sim 2.5\times 10^{-16}$, which gives
\beqn
{\cal B}(\pi^0\to \bar{\nu}\nu \bar{\nu}\nu)\sim 4 \times 10^{-23},\\
{\cal B}(\eta\to \bar{\nu}\nu \bar{\nu}\nu)\sim 1 \times 10^{-18},\\
{\cal B}(\eta^\prime\to \bar{\nu}\nu \bar{\nu}\nu)\sim 5 \times 10^{-18},
\eeqn
by using the experimental data of $P\to\gamma\gamma$ \cite{PDG2018}. Although these branching ratios are still small and far from the present experimental bounds, comparing with eq. (\ref{br-pinunu}) for one neutrino pair, one can find that there is a large enhancement for the four-neutrino final states, and the enhancement could be increasing when the decaying pseudoscalar particle gets heavier. Therefore, for heavy pseudo-quarkonia like $\eta_c$ or $\eta_b$, a much stronger enhancement may be expected. Similar enhancement has been pointed out in the neutral $B$ and $D$ invisible decays by the authors of Ref. \cite{BGP2018}.

To summarize, we have calculated the lowest order SM contribution to the branching ratios of the light vector meson invisible decays, ${\cal B}(V\to\bar{\nu}\nu)$, for $V=\rho,\;\omega,\;\phi$. Our theoretical results for ${\cal B}(V\to\bar{\nu}\nu)$ are about $10^{-13}\sim 10^{-11}$, which are far below the upper limits given by the BESIII experiment very recently. The $J/\psi\to \eta(\eta^\prime)\bar{\nu}\nu$ decays have been studied, and their decay rates were estimated. We also revisit the light pseudoscalar invisible decays. It is known that $P\to \bar{\nu}\nu$ is strongly helicity-suppressed; however, such suppression does not happen in $P\to\bar{\nu}\nu \bar{\nu}\nu$.  These results may provide some complementary information for the future experimental and theoretical investigations of the invisible decays of light mesons, in order to explore new physics scenarios beyond the SM.

\vspace{0.5cm}
\section*{Acknowledgements}
This work was supported in part by the NSF of China under Grant No. 11575175, and by the CAS Center for Excellence in Particle Physics (CCEPP).


\begin{thebibliography}{40}
\bibitem{BESII2008}M. Ablikim {\it et al.} (BESII Collaboration), Phys. Rev. Lett. {\bf 100}, 192001 (2008), arXiv:0710.0039 [hep-ex].
\bibitem{Babar2009}B. Aubert  {\it et al.} ($BABAR$ Collaboration), Phys. Rev. Lett. {\bf 103}, 251801 (2009), arXiv:0908.2840 [hep-ex].
\bibitem{CLN1998}L.N. Chang, O. Lebedev, and J.N. Ng, Phys. Lett. B {\bf 441}, 419 (1998), hep-ph/9806487.
\bibitem{E9492005}A.V. Artamonov {\it et al.} (E949 Collaboration), Phys. Rev. D {\bf 72}, 091102 (2005), hep-ex/0506028.
\bibitem{BESIII2013}M. Ablikim {\it et al.} (BESIII Collaboration), Phys. Rev. D {\bf 87}, 012009 (2013), arXiv:1209.2469 [hep-ex].
\bibitem{AMP1982}L. Arnellos, W.J. Marciano, and Z. Parsa,  Nucl. Phys. {\bf B196}, 365 (1982).
\bibitem{BESIII2018}M. Ablikim  {\it et al.} (BESIII Collaboration), Phys. Rev. D {\bf 98}, 032001 (2018), arXiv:1809.05613 [hep-ex].
\bibitem{PDG2018}M. Tanabashi {\it et al.} (Particle Data Group), Phys. Rev. D {\bf 98}, 030001 (2018).
\bibitem{McElrath2007}B. McElrath, {\it Light Higgses and Dark Matter at Bottom and Charm Factories}, eConf C070805, 19 (2007), arXiv:0712.0016 [hep-ph].
\bibitem{BP2010}A. Badin and A.A. Petrov, Phys. Rev. D {\bf 82}, 034005 (2010), arXiv:1005.1277 [hep-ph].
\bibitem{BFG2018}D. Barducci, M. Fabbrichesi, and E. Gabrielli, Phys.Rev. D {\bf 98}, 035049 (2018), arXiv:1806.05678 [hep-ph].
\bibitem{NSVZ1980}V.A. Novikov, M.A. Shifman, A.I. Vainshtein, and V.I. Zakharov, Nucl. Phys. {\bf B165}, 55 (1980).
\bibitem{KKKS1983}J.G. K\"orner, J.H. K\"uhn, M. Krammer, and H. Schneider, Nucl. Phys. {\bf B229}, 115 (1983).
\bibitem{YLSZ1984}H. Yu, B.A. Li, Q.X. Shen, and M.M. Zhang, High Energy Phys. \& Nucl. Phys. {\bf 8}, 285 (1984) (in Chinese).
\bibitem{Holdom1986}B. Holdom, Phys. Lett. {\bf 166B}, 196 (1986).
\bibitem{BESIII2018-2}M. Ablikim {\it et al.} (BESIII Collaboration), arXiv:1809.00635 [hep-ex]; arXiv:1810.03091 [hep-ex].
\bibitem{Liyang2012}J. Fu, H.B. Li, X. Qin, and M.Z. Yang,  Mod. Phys. Lett. A {\bf 27}, 1250223 (2012), arXiv:1111.4055 [hep-ph].
\bibitem{Gninenko2015}S.N. Gninenko,  Phys. Rev. D {\bf 91}, 015004 (2015),  arXiv:1409.2288 [hep-ph].
\bibitem{GHMT2006}A. Goobar, S. Hannestad, E. Mortsell, and H. Tu, JCAP {\bf 0606}, 019 (2006), astro-ph/0602155.
\bibitem{BGP2018}B. Bhattacharya, C.M. Grant, and A.A. Petrov, arXiv:1809.04606 [hep-ph].

\end{thebibliography}
\end{document}